\documentclass[aps,pra,preprint,groupedaddress,showpacs]{revtex4}
\usepackage{bm}
\usepackage{epsf}
\usepackage{amssymb}
\usepackage{amsmath}
\usepackage{graphicx}
\usepackage{rotating}
\usepackage{epsfig}
\usepackage{psfrag}
\usepackage{amsmath}
\usepackage{hyperref}

\hypersetup{
    bookmarks=true,         
    unicode=false,          
    pdftoolbar=true,        
    pdfmenubar=true,        
    pdffitwindow=true,      
    pdftitle={My title},    
    pdfauthor={Author},     
    pdfsubject={Subject},   
    pdfcreator={Creator},   
    pdfproducer={Producer}, 
    pdfkeywords={keywords}, 
    pdfnewwindow=true,      
    colorlinks=false,       
    linkcolor=red,          
    citecolor=green,        
    filecolor=magenta,      
    urlcolor=cyan           
}

\DeclareMathAlphabet{\bi}{OML}{cmm}{b}{it}

\begin{document}

\def\ba{\mathbf{a}}
\def\d{\mathbf{d}}
\def\P{\mathbf{P}}
\def\bP{\mathbf{P}}
\def\bF{\mathbf{F}}
\def\bk{\mathbf{k}}
\def\bkn{\mathbf{k}_{0}}
\def\bx{\mathbf{x}}
\def\bfn{\mathbf{f}}
\def\bg{\mathbf{g}}
\def\bj{\mathbf{j}}
\def\bR{\mathbf{R}}
\def\bfr{\mathbf{r}}
\def\bu{\mathbf{u}}
\def\bq{\mathbf{q}}
\def\bw{\mathbf{w}}
\def\bp{\mathbf{p}}
\def\bG{\mathbf{G}}
\def\bz{\mathbf{z}}
\def\bs{\mathbf{s}}
\def\E{\mathbf{E}}
\def\bv{\mathbf{v}}
\def\b0{\mathbf{0}}
\def\la{\langle}
\def\ra{\rangle}
\def\beq{\begin{equation}}
\def\eeq{\end{equation}}
\def\bea{\begin{eqnarray}}
\def\eea{\end{eqnarray}}
\def\bnab{\bm{\nabla}}
\def\bA{{\bf A}}
\def\R{R^0_{\mathrm{TF}}}
\def\L{\bm{\Lambda}}
\def\bua{{\mathbf{u}}^{\mathrm{ad}}}
\def\bus{{\mathbf{u}}^{\mathrm{en}}}
\def\lp{\epsilon_{\mathrm{LP}}}

\title{Second sound and the density response function in uniform superfluid atomic gases}
\author{H.~Hu}
\affiliation{ACQAO and Centre for Atom Optics and Ultrafast Spectroscopy, Swinburne University of Technology, Melbourne, Victoria 3122, Australia}
\affiliation{Department of Physics, Renmin University of China, Beijing 100872, China}
\author{E.~Taylor}
\affiliation{Department of Physics, The Ohio State University, Columbus, Ohio, 43210, USA}
\author{X.-J.~Liu}
\affiliation{ACQAO and Centre for Atom Optics and Ultrafast Spectroscopy, Swinburne University of Technology, Melbourne, Victoria 3122, Australia}
\author{S.~Stringari}
\affiliation{CNR-INFM BEC Centre and Dipartimento~di~Fisica, Universit\`a di Trento, I-38050 Povo, Trento, Italy}
\author{A.~Griffin}
\affiliation{Department of Physics, University of Toronto, Toronto, Ontario, M5S 1A7, 
Canada}

\date{Apr. 26, 2010}

\begin{abstract}
Recently there has been renewed interest in second sound in superfluid Bose and Fermi gases. By using two-fluid hydrodynamic theory, we review the density response $\chi_{nn}(\bq,\omega)$ of these systems as a tool to identify second sound in experiments based on density probes.  Our work generalizes the well-known studies of the dynamic structure factor $S(\bq,\omega)$ in superfluid $^4$He in the critical region.  
We show that, in the unitary limit of uniform superfluid Fermi gases, the relative weight of second vs. first sound in the compressibility sum rule is given by the Landau--Placzek ratio $\lp\equiv (\bar{c}_p-\bar{c}_v)/\bar{c}_v$ for all temperatures below $T_c$.  In contrast to superfluid $^4$He, $\lp$ is much larger in strongly interacting Fermi gases, being already of order unity for $T \sim 0.8 T_c$, thereby providing promising opportunities to excite second sound with density probes. The relative weights of first and second sound are quite different in $S(\bq,\omega)$ (measured in pulse propagation studies) as compared to $\mathrm{Im}\chi_{nn}(\bq,\omega)$ (measured in two-photon Bragg scattering).  We show that first and second sound in $S(\bq,\omega)$ in a strongly interacting Bose-condensed gas are similar to those in a Fermi gas at unitarity.  However, in a weakly interacting Bose gas, first and second sound are mainly uncoupled oscillations of the thermal cloud and condensate, respectively, and second sound has most of the spectral weight in $S(\bq,\omega)$.  We also discuss the behaviour of the superfluid and normal fluid velocity fields involved in first and second sound.
\end{abstract}

\maketitle

\section{Introduction}

The most dramatic effects related to superfluidity in liquid $^4$He arise~\cite{clarendon67} when the dynamics of the two components are described by the two-fluid hydrodynamics first discussed by Landau~\cite{cambridge41}. These equations only describe the dynamics when the non-equilibrium states are in local hydrodynamic equilibrium~\cite{cambridge09}, which requires short collision times between the excitations forming the normal fluid. (This requirement is usually summarized as $\omega\tau\ll 1$, where $\omega$ is the frequency of a collective mode and $\tau$ is the appropriate relaxation rate.) The study of ultracold gases when they are in local equilibrium has been difficult because the density and the \textit{s}-wave scattering length are typically not large enough. However, recent experimental work on trapped Bose-condensed gases has reported some success, with evidence for a second sound mode in highly elongated (cigar-shaped) traps~\cite{meppelink09}.

Another approach to achieving conditions where the Landau two-fluid description is correct has been to consider a Fermi superfluid gas close to unitarity~\cite{giorpitastrin08}, where the \textit{s}-wave scattering length between Fermi atoms in two different hyperfine states is infinite. Developing earlier theoretical studies~\cite{taygrif05,tahugrif08}, Taylor and co-workers~\cite{tayhu09} have recently given detailed predictions for the first and second sound breathing oscillations in a trapped Fermi gas at unitarity.
The coupled differential equations of the Landau two-fluid description were solved variationally, with results which agreed with analytic predictions at $T\to 0$ and $T\to T_c$. So far, only an isotropic trap has been studied.

In \cite{tayhu09}, it was shown that (as in superfluid $^4$He) the frequencies of first and second sound in unitary Fermi gases were quite well approximated by assuming that the solutions of the two-fluid equations corresponded to pure uncoupled density and temperature waves.  This is in sharp contrast to the situation in dilute, weakly interacting  Bose gases, where first and second sound involve both density and temperature oscillations~\cite{ZNGJLTP}.  Na\"ively, the first result above would lead one to believe that second sound in a unitary Fermi gas would have very small weight in the density response function. However, Arahata and Nikuni~\cite{arahatanik} have shown that in a uniform Fermi gas at unitarity, a density disturbance can have first and second sound pulses of comparable magnitude at temperatures of order $0.8 T_c$. 

The purpose of this paper is to provide a systematic study of the density response function $\chi_{nn}(\bq,\omega)$ of a uniform superfluid atomic gas in the hydrodynamic regime~\cite{hohmar64,Hohenberg65}, as described by the non-dissipative Landau two-fluid equations (see sections~\ref{sqomegasec} and \ref{sqomegatwofluidsec}).  In this hydrodynamic region, the related dynamic structure factor is given by $S(\bq,\omega)\propto \mathrm{Im}\chi_{nn}(\bq,\omega)/\omega$.  In studies of superfluid $^4$He, $S(\bq,\omega)$ is directly measured in Brillouin light scattering experiments~\cite{greytak,vinen70}.  In dilute gases, density pulse propagation experiments effectively measure $S(\bq,\omega)$ as noted in \cite{arahatanik}.  In addition, experiments using two-photon Bragg scattering have a cross-section proportional to $\mathrm{Im}\chi_{nn}(\bq,\omega)$.  While all these experiments probe the same density response function, $\mathrm{Im}\chi_{nn}(\bq,\omega)$, the relative weight of the low frequency resonances are strongest in $S(\bq,\omega)$ because of the extra factor of $1/\omega$ noted above.  This fact has significant implications for measuring second sound since the frequency of second sound becomes much smaller than first sound as $T_c$ is approached.  

We relate our analysis to classic discussions of the dynamic structure factor $S(\bq,\omega)$ in the two-fluid region of superfluid $^4$He in the critical region~\cite{greytak,vinen70,Hohenberg73,winterling73,connor75}.  However, many new features arise when dealing with quantum gases.

New aspects of our work include the use of the compressibility sum rule (sections~\ref{sqomegatwofluidsec} and \ref{sqomegaweightssec}) in understanding the relative weights of first and second sound in the density response function.  
We also discuss the role played by the Landau--Placzek ratio $\lp \equiv (\bar{c}_p-\bar{c}_v)/\bar{c}_v$ in determining these weights, with emphasis on new aspects that arise in superfluid gases.
Here, $\bar{c}_p$ and $\bar{c}_v$ are the specific heats per unit mass at constant pressure and volume, respectively.  We discuss the differences between $S(\bq,\omega)$ of a unitary Fermi gas and that of a Bose-condensed gas (sections~\ref{sqomegaweightssec} and \ref{Bosegassec}), and present results for Bragg scattering with localized beams applied to the centre of a trapped unitary Fermi gas using the local density approximation (section~\ref{Braggsec}). In~\ref{vfieldssec}, we provide a detailed analysis of the superfluid and normal fluid velocity fields associated with first and second sound.

\section{Density response function}
\label{sqomegasec}

In this section, we review some basic properties of the density response function $\chi_{nn}(\bq,\omega)$, the related dynamic structure factor $S(\bq,\omega)$, and the experiments that measure these quantities.  The density response function involves the correlation between the density at two different points at different times:
\beq \chi_{nn} (\bfr, t)\equiv -i\theta(t)  \left\langle\left[{\hat\rho}(\bfr, t), {\hat\rho}(\b0, 0)\right]\right\rangle.\label{chinn}\eeq
Here, $\theta(t)$ is the step function and we have set the volume equal to unity.  The Fourier transform $\mathrm{Im}\chi_{nn}(\bq,\omega)$ of the imaginary part of the density response function is related to the dynamic structure factor $S(\bq,\omega)$ by
\bea {\rm Im}\,\chi_{nn}(\bq, \omega + i0^+) &=& - n\pi [S(\bq, \omega) - S(-\bq, -\omega)] \nonumber\\
&=& -n\pi\, S(\bq, \omega) (1-e^{-\beta\hbar\omega}),\label{Imchinn}\eea
where $n$ is the density, $\beta = 1/T$ is the inverse temperature, and in the last step we have made use of the ``detailed balance" relation, $S(\bq, -\omega) = e^{-\beta\omega} S(\bq, \omega)$ and also the inversion symmetry of the system which gives $S(-\bq,\omega)=S(\bq,\omega)$.  Here and throughout this paper we set $\hbar=k_B=1$.  Equation~(\ref{Imchinn}) can also be re-written as
\beq S(\bq, \omega) = - {1\over \pi n} [N^0(\omega) + 1]{\rm Im}\, \chi_{nn}(\bq, \omega + i0^+),\label{sqomega2}\eeq
where $N^0(\omega) = (e^{\beta\omega} - 1)^{-1}$ is the Bose distribution function.  The latter arises in both Bose and Fermi fluids because density fluctuations always obey Bose statistics.  We see from (\ref{Imchinn}) that $\mathrm{Im}\chi_{nn}(\bq,\omega)$ is the antisymmetric (frequency) part of the dynamic structure factor $S(\bq,\omega)$.

The dynamic structure factor $S(\bq,\omega)$ is measured by using inelastic scattering (e.g., neutron and Brillouin light scattering) of probe particles, which transfer momentum $\bq$ and energy $\omega$ to the system.  $S(\bq,\omega)$ is discussed and calculated in all standard texts on many body physics (convenient reviews are given in \cite{TrentoBook,griffin93}).
In contrast, the density response function $\chi_{nn}(\bq,\omega)$ describes the density fluctuation induced by an external perturbing potential.  As an example, in atomic gases, the imaginary part of the density response function is probed in two-photon Bragg scattering and also describes the density pulse produced by a blue-detuned laser.  Brillouin and neutron scattering have been used extensively to study the collective modes in $^4$He.  These techniques cannot be applied to dilute atomic gases, however, since the gases are far too dilute to generate an appreciable signal.  Instead, spectroscopic probes of dilute gases have successfully used Bragg scattering since this technique makes use of the stimulated light scattering processes induced by two counter-propagating laser beams, resulting in a strong enhancement of the signal.

In Bragg scattering experiments (see, e.g.,~\cite{StamperKurn99}), by measuring the total momentum transferred to the sample, one measures  the imaginary part of the density response function $\chi_{nn}(\bq,\omega)$, where $\bq$ and $\omega$ are the momentum and energy transferred by the stimulated absorption and emission of the photons~\cite{TrentoBook}.  If the Bragg pulse is short compared to $2\pi/\omega_z$, where $\omega_z$ is the frequency of the harmonic trap along the axis of light propagation ($\bq=\hat{\bz}q$), the momentum transferred is
\beq \Delta P_z = 2q\tau\left(\frac{V}{2}\right)^2\mathrm{Im}\chi_{nn}(q,\omega),\label{DeltaPuniform}\eeq
where $V$ is the strength of the potential induced by the lasers and $\tau$ is the pulse duration.

\begin{table*}
\begin{center}
\renewcommand{\arraystretch}{2}
\begin{tabular}[c]{|c|c|c|c|}
\hline
\multicolumn{1}{|c|}{Method}&
\multicolumn{1}{c|}{System}&
\multicolumn{1}{c|}{Quantity measured}&
\multicolumn{1}{c|}{Response function}
\\ \hline
Bragg Spectroscopy & Gases & Momentum transferred & $\mathrm{Im}\chi_{nn}(\bq,\omega)$
\\ \hline
Neutron Spectroscopy & $^4$He & Number of scattered neutrons & $S(\bq,\omega)$
 \\ \hline
Brillouin Spectroscopy
($\omega \ll T$) & $^4$He & Number of scattered photons & $S(\bq,\omega) \sim \frac{T }{\omega}\mathrm{Im}\chi_{nn}(\bq,\omega)$
\\ \hline
Density pulse & Gases & Pulse amplitude & $\frac{1}{\omega}\mathrm{Im}\chi_{nn}(\bq,\omega)$
\\ \hline
\end{tabular}
\caption{Summary of experimental probes.  From left to right: the experimental method, the system(s) to which the method can be applied, the quantity being measured, and the density response function involved.  In the two-fluid region of strongly interacting Fermi gases and $^4$He, second sound is weakly coupled to density fluctuations but has a small frequency $\omega$ at high temperatures (below $T_c$).  This means that, as a result of the extra factor of $1/\omega$ multiplying the density response function,  Brillouin spectroscopy and density pulses are more sensitive to second sound than Bragg and neutron scattering.}
\label{probes}
\end{center}
\end{table*}

As noted recently by Arahata and Nikuni~\cite{arahatanik}, the density response function can also be measured by exciting a density pulse in a uniform gas. Within linear response theory, the density fluctuation $\delta n(\bfr,t)$ induced by the application of an external perturbing potential $\delta V(\bfr,t)$ is
\beq
\delta n(\bfr,t)=\int \frac{ d\mathbf{q}}{(2\pi)^3} \int \frac{d\omega}{2\pi} \chi_{nn}(\bq,\omega)\delta V(\bq,\omega)e^{i\mathbf{q}\cdot \mathbf{r}-i\omega t},
\eeq
where $\delta V(\bq, \omega)$ is the Fourier transform of the perturbing potential. Similar to the sound propagation experiments in \cite{Andrews97,Joseph07}, consider a localized potential applied for a short duration $\tau<t<0$, and turned off at $t=0$. Assuming for simplicity that the perturbing potential only varies spatially along the $z$-axis (and $\bq= \hat{\bz}q$), for $t>0$, one obtains~\cite{arahatanik}
\beq
\delta n(z,t)=\frac{1}{2\pi^2}\int  dq\int d\omega \delta V(q)
\frac{\mathrm{Im}\chi_{nn}(q,\omega)}{\omega}e^{iqz-i\omega t}.\label{dfluct2}
\eeq 
Equation~(\ref{dfluct2}) reveals an important feature: in contrast to Bragg scattering, where the response is proportional to $\mathrm{Im}\chi_{nn}(\bq,\omega)$, the response involved in the excitation of density pulses in proportional to $\mathrm{Im}\chi_{nn}(\bq,\omega)/\omega$.  As we discuss in section~\ref{sqomegatwofluidsec}, for uniform superfluids, this factor of $1/\omega$ leads to an enhancement in the signal of second sound in density pulse experiments as compared to its signal in Bragg spectroscopy.  In table~\ref{probes}, we review the quantities measured by the experimental probes of interest in both dilute gases and superfluid $^4$He.  

Before discussing the form of the density response function in the two-fluid hydrodynamic region in section~\ref{sqomegatwofluidsec}, we first review below some general properties of the dynamic structure factor that will be of use later on in understanding the contributions from first and second sound in experimental probes of superfluid atomic gases.

Quite generally, $S(\bq, \omega)$ satisfies various frequency moment sum rules.  Two are of special particular interest in the two-fluid domain~\cite{nozierespines90}.  The $f$-sum rule (valid for all $q$) is
\beq \int^\infty_{-\infty} \omega\, S(\bq, \omega) = {q^2\over 2m}. \label{fsumrule}\eeq
The compressibility sum rule arises from the exact Kramers-Kronig identity (also valid for all $q$)
\beq \chi_{nn} (\bq, \omega = 0)=\int^\infty_{-\infty} {d\omega'\over\pi} \frac{\mathrm{Im}\chi_{nn}(\bq, \omega')}{\omega'},\label{KK}\eeq
which involves an inverse frequency moment.  (Note that $\chi_{nn}(\bq,\omega=0)=\mathrm{Re}\chi_{nn}(\bq,\omega=0)$ is purely real.)  Using (\ref{Imchinn}), one can show that (\ref{KK}) is equivalent to
\beq \chi_{nn}(\bq, \omega=0) = -2n \int^\infty_{-\infty}d\omega' {S(\bq, \omega')\over\omega'}.\label{chinnidentity}\eeq
The usefulness of (\ref{KK}) and (\ref{chinnidentity}) is due to the fact that in the long wavelength limit $(\bq \to 0),$ the density response function $\chi_{nn}(\bq = 0, \omega=0)$ describes static thermodynamic fluctuations which can be related to the equilibrium isothermal  compressibility~\cite{nozierespines90,mazenko06}
\beq \lim_{q\to 0}\chi_{nn} (\bq, \omega =0)=-n\left.{\partial n \over\partial p}\right|_{T}.\label{compressibility}\eeq
Thus we arrive at the so-called compressibility sum rule for the dynamic structure factor,
\beq \lim_{q\to 0} \int^\infty_{-\infty}d\omega' {S(\bq, \omega')\over \omega'} = {1\over 2m} \left.{\partial \rho \over\partial p}\right|_{T}\equiv  {1\over 2m v^2_T},\label{compressibilitysumrule}\eeq
where we have introduced the isothermal sound velocity $v_T$.

The sum rules in (\ref{fsumrule}) and (\ref{compressibilitysumrule}) are general and may be viewed as two constraints which $S(\bq, \omega)$ must always satisfy.  The $f-$sum rule is used frequently in discussions of inelastic neutron scattering as a check on the experimental data.  As we discuss in section~\ref{sqomegatwofluidsec}, the compressibility sum rule has special significance when we discuss the two-fluid hydrodynamic region, as first emphasized by Nozi\`{e}res and Pines~\cite{nozierespines90}.

\section{Dynamic structure factor in the two-fluid regime}
\label{sqomegatwofluidsec}

The Landau two-fluid equations for an isotropic (such that $\chi_{nn}(\bq,\omega)=\chi_{nn}(q,\omega)$) superfluid (Bose or Fermi) describe the dynamics when collisions are strong enough to produce local equilibrium.  Using the non-dissipative two-fluid equations, one finds the density response function (see Refs.~\cite{hohmar64,Hohenberg65} and page~138 of \cite{cambridge09}), 
\beq \chi_{nn}(q, \omega) = {nq^2\over m} {\omega^2-v^2q^2\over (\omega^2 - u^2_1q^2)(\omega^2-u^2_2q^2)},\label{chinntwofluid}\eeq
where we have defined a new velocity
\beq v^2\equiv T{\bar{s}^2_0\over {\bar c}_v} {\rho_{s0}\over\rho_{n0}}.\label{vspeed}\eeq
Here, $\bar{s}_0\equiv S_0/Nm$ is the equilibrium entropy $S_0$ per unit mass and $\bar c_v \equiv T(\partial \bar s/\partial T)_{\rho}$ is the specific heat per unit mass at constant volume.  The equilibrium superfluid and normal fluid densities are denoted by $\rho_{s0}$ and $\rho_{n0}$, respectively.  In (\ref{chinntwofluid}), $u_1$ and $u_2$ are the well-known exact first and second sound velocities given by Landau's hydrodynamics.  For later purposes, we note that these velocities satisfy the exact relations
\beq u^2_1 + u^2_2 = v^2 +v^2_{\bar{s}},\;\;\; u^2_1 u^2_2 = v^2_Tv^2 = v^2_{\bar{s}}\frac{v^2}{\gamma},\label{speedidentities}\eeq
where $v_T$ is defined in (\ref{compressibilitysumrule}) and the adiabatic sound velocity is defined as
\beq v^2_{\bar{s}} \equiv\left.{\partial P\over\partial\rho}\right|_{\bar s}.\label{adiabaticspeed}\eeq
We note that thermodynamic relations~\cite{mazenko06} show that, quite generally, the ratio of the specific heats can be related to the isothermal and adiabatic sound velocities,
\beq \gamma\equiv{\bar c_p\over \bar c_v} = {v^2_{\bar{s}}\over v^2_T}. \label{gammadef}\eeq
The difference between $v_{\bar{s}}$ and $v_T$ is a consequence of the finite thermal expansion.  

Equation~(\ref{chinntwofluid}) gives the density response for Bose and Fermi fluids described by the non-dissipative Landau two-fluid equations.  The two-fluid density response function in the presence of dissipation arising from transport coefficients has been given by Hohenberg and Martin~\cite{hohmar64,Hohenberg65,Hohenberg73}.  (See also the discussion given by Vinen~\cite{vinen70}.) For our purposes in this paper, the primary effect of dissipation is to broaden the delta function peaks in the imaginary part of the non-dissipative density response function.  It does not otherwise change the structure of the two-fluid response function and in particular, does not affect in a significant way the weights of first and second sound in this function.    

From (\ref{chinntwofluid}), it is straightforward to obtain the imaginary part of the density response function, which is the experimentally relevant quantity in dilute gases:
\bea \mathrm{Im}\chi_{nn}(q,\omega) &=& -n\pi\frac{q^2}{2m}\bigg\{\!{Z_1\over \omega}\left[\delta(\omega-u_1q)\!+\!\delta(\omega + u_1q)\right]\nonumber\\
&&+{Z_2\over\omega}\left[\delta(\omega-u_2q)+\delta(\omega + u_2q)\right]\!\bigg\}.\label{Imchinntwofluid}\eea
Here, we have defined the ``traditional" amplitudes for first and second sound, namely
\beq Z_1 \equiv {u^2_1 - v^2\over u^2_1 - u^2_2}\ , \ Z_2 \equiv {v^2-u^2_2\over u^2_1 - u^2_2}.\label{Z1Z2def}\eeq
Using (\ref{Imchinntwofluid}) in (\ref{sqomega2}), the two-fluid expression for the dynamic structure factor is found to be
\bea S(q, \omega)\!\! &=& \!\!{q^2\over 2m} [N^0(\omega)+1]\bigg\{\!{Z_1\over \omega}\left[\delta(\omega-u_1q)\!+\!\delta(\omega + u_1q)\right]\nonumber\\
&&+{Z_2\over\omega}\left[\delta(\omega-u_2q)+\delta(\omega + u_2q)\right]\!\bigg\}.\label{sqomegatwofluid}\eea

The two-fluid dynamic structure factor in (\ref{sqomegatwofluid}) turns out to satisfy both the $f$-sum and compressibility sum rules, given by (\ref{fsumrule}) and (\ref{compressibilitysumrule})~\cite{nozierespines90}.  This makes use of the identity $N^0 (\omega) + 1+N^0 (-\omega)+1 = 1$, satisfied by the Bose distribution.  
Substituting (\ref{sqomegatwofluid}) into the left-hand side of (\ref{fsumrule}), one sees that the amplitudes $Z_1$ and $Z_2$ describe the weights of first and second sound, respectively, in the $f$-sum rule.  That the sum of these two contributions saturate the $f$-sum rule is a consequence of the fact that $Z_1+Z_2 = 1$. 
Similarly, substituting (\ref{sqomegatwofluid}) into the left-hand side of (\ref{compressibilitysumrule}), one sees the relative contributions of first and second sound to the isothermal compressibility are given by $Z_1/u^2_1$ and $Z_2/u^2_2$, respectively.  We will denote these weights by $W_i$:
\beq W_1 \equiv \frac{Z_1}{u^2_1},\;\;W_2\equiv \frac{Z_2}{u^2_2}.\label{compressibilityweights} \eeq
To see that the compressibility sum rule is saturated by these contributions from first and second sound, we note that
$Z_1/u^2_1+Z_2/u^2_2 =1/v^2_T$, as can be seen from (\ref{speedidentities}).

Since we are working in the low frequency two-fluid hydrodynamic region, we can use the fact that $\omega\ll  T$. In this limit, the detailed balance factor in (\ref{sqomegatwofluid}) can be approximated by
\beq N^0(\omega)+1\simeq {T\over\omega} +{1\over 2} +\dots \label{hydrolimit}\eeq
The first quantum correction, given by the $1/2$ term in (\ref{hydrolimit}) in this expansion, is needed to satisfy the sum rules previously discussed. This approximation $(\omega\ll T)$ is often referred to as the ``classical limit'' in textbooks discussing the dynamic structure factor, although in fact it describes both superfluid ($T<T_c$) and normal fluid ($T>T_c$) collisional hydrodynamics.

Keeping only the leading order first term in (\ref{hydrolimit}), (\ref{sqomegatwofluid}) simplifies to
\bea S(q, \omega) &\simeq& {T\over 2m}\! \bigg\{{Z_1\over u^2_1} \left[\delta(\omega - u_1 q)\!+\!\delta(\omega + u_1 q)\right]\nonumber\\
&&+{Z_2\over u^2_2} \left[\delta(\omega-u_2q)\!+\!\delta(\omega +u_2q)\right]\bigg\}\nonumber\\
&=&-\frac{1}{\pi n}\left(\frac{T}{\omega}\right)\mathrm{Im}\chi_{nn}(q,\omega).\label{sqomegahydro}\eea
This expression is valid for Bose and Fermi superfluid gases as well as superfluid $^4$He, that is, all superfluids described by an order parameter with a phase and amplitude.
Equation~(\ref{sqomegahydro}) shows that in the low-frequency two-fluid hydrodynamic region, $W_i\equiv Z_i/u^2_i$ gives the weight of the $i^{\rm th}$ mode.  In contrast, the weight of the $i^{\rm th}$ mode in Bragg scattering [see (\ref{DeltaPuniform}) and (\ref{Imchinntwofluid})] is given by $B_i$, where
\beq B_1 \equiv \frac{Z_1}{u_1},\;\;B_2\equiv \frac{Z_2}{u_2}.\label{Braggweights} \eeq
Comparing (\ref{sqomegahydro}) with (\ref{Imchinntwofluid}), we see that the hydrodynamic limit ($\omega\ll T$) of $S(q,\omega)$ involves an extra factor of $1/\omega$.  

We see that while $\mathrm{Im}\chi_{nn}(q,\omega)$ and $\mathrm{Im}\chi_{nn}(q,\omega)/\omega$ share the same poles, the relative weights of first and second sound are quite different in these two functions, with $B_2/B_1 = (W_2/W_1)(u_2/u_1)$ generally being much smaller than $W_2/W_1$ due to the smallness of $u_2/u_1$ (see Fig.~\ref{soundfig}).

The two-fluid expression in (\ref{sqomegahydro}) is known from earlier discussions of Brillouin scattering in superfluid $^4$He (see section~\ref{sqomegaweightssec}). In contrast to high-frequency inelastic neutron scattering, Brillouin scattering is carried out with light in the visible part of the spectrum ($\omega \ll T$).  As pointed out above [see (\ref{sqomegahydro})], this means that Brillouin scattering measures the imaginary part of the density response function divided by the frequency $\omega$.  In studies of superfluid $^4$He, this factor of $1/\omega$ is crucial to probing second sound since, although only weakly coupled to density, the speed of second sound is small and hence, its weight in Brillouin light scattering can be significant.  

As noted in section~\ref{sqomegasec}, a similar situation arises in studies of pulse propagation in dilute gases, albeit for very different reasons.   In this situation [see (\ref{dfluct2})], the amplitude of the density pulse induced by a localized (in time and space) density perturbation is proportional to $\mathrm{Im}\chi_{nn}(q,\omega)/\omega$.  The factor of $1/\omega$ here results not from the detailed balance factor, but from the Fourier transform of the step-function used to model the time dependence of the perturbing laser beams.  Arahata and Nikuni~\cite{arahatanik} calculated the effect of producing a density fluctuation by a sudden perturbation in a superfluid Fermi gas at unitarity.  The final result was two pulses moving with the speeds of first and second sound, with relative amplitudes given by $W_i$ as defined in (\ref{compressibilityweights}).

\section{Superfluids with small thermal expansion}
\label{sqomegaweightssec}

In section~\ref{sqomegatwofluidsec}, we showed that the relative weights in $S(q, \omega)$ and $\mathrm{Im}\chi_{nn}(q,\omega)$ of first and second sound are given by $W_i=Z_i/u^2_i$ and $B_i=Z_i/u_i$, respectively, where $Z_i$ is defined in (\ref{Z1Z2def}).  These results are valid for any type of superfluid, including weakly interacting atomic Bose gases and also strongly interacting Fermi gases.  In this section, we concentrate on strongly interacting Fermi gases where, as discussed above, the thermal expansion is relatively small, meaning that density and temperature fluctuations are not strongly coupled.   Using the exact relations in (\ref{speedidentities}), in this section we will show that to a very good approximation, $W_2/W_1$ is given by (\ref{W2overW1}) in this situation.  In section~\ref{Bosegassec}, we give a brief discussion of the dynamic structure factor in a dilute Bose gas, where the thermal expansion can be much larger when the $s$-wave scattering length is small.

\begin{figure}
\begin{center}
\epsfig{file=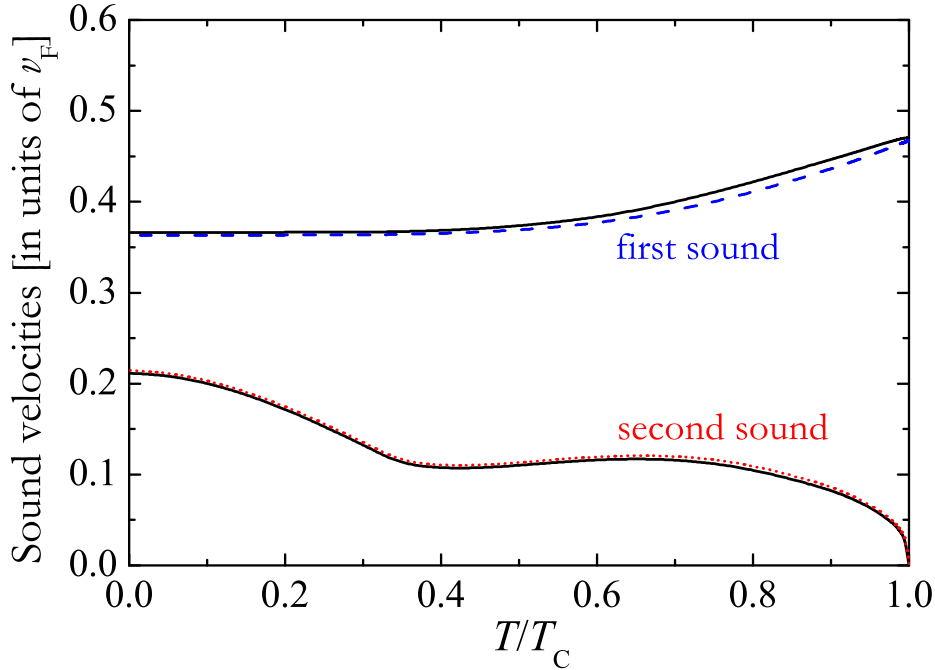, angle=0, width=0.75\textwidth}
\caption{ Speeds of first and second sound in a unitary superfluid Fermi gas.  The dashed lines are the approximations $c_1 = v_{\bar{s}}$ and $c_2 = v/\sqrt{\gamma}$, as given by the leading terms in (\ref{uexpansion}) (from \cite{tayhu09}).}
\label{soundfig}
\end{center}
\end{figure}

Solving the coupled equations in (\ref{speedidentities}), one obtains the expansions~\cite{vinen70,Hohenberg73}
\beq u^2_1 = v^2_{\bar{s}}[1 + (\gamma-1)x + \cdots],\;\;
u^2_2 = \frac{v^2}{\gamma}[1-(\gamma-1)x+\cdots],\label{uexpansion}\eeq
in terms of the parameter defined by $x\equiv v^2/\gamma v^2_{\bar{s}}$.  These results show that to lowest order in the parameter $x(\gamma-1)\equiv x\lp$, the first and second sound velocities are well approximated by
\beq c^2_1 = v^2_{\bar{s}},\;\;\;c^2_2 = \frac{v^2}{\gamma}= T\frac{\bar{s}^2_0}{\bar{c}_p}\frac{\rho_{s0}}{\rho_{n0}}.\label{cspeeds}\eeq
In figure~\ref{soundfig}, we show the temperature dependence of $u_1$ and $u_2$ and compare these results with the leading order expressions $c_1$ and $c_2$ given by (\ref{cspeeds}) (shown by the dashed lines).  The thermodynamic functions needed to obtain these sound speeds are calculated using the microscopic theory of Nozi\`eres and Schmitt-Rink (NSR)~\cite{NSR,Hu06}.  We see that $c_1$ and $c_2$ are extremely good approximations at all temperatures.  At temperatures $T\gtrsim 0.4T_c$, $x=c^2_2/c^2_1$ is very small, even though $\lp\sim{\cal{O}}(1)$. At lower temperatures, while $x$ is no longer small ($x=1/3$ at $T=0$), $\lp$ becomes extremely small.  Thus in both limits, we find that the correction term $x\lp$ in (\ref{uexpansion}) is negligible.  

The fact that $u_1\simeq c_1$ and $u_2\simeq c_2$ in a fluid with small thermal expansion such as the unitary Fermi gas means that first and second sound propagate at essentially the same velocities as pure density and temperature oscillations, as would arise when $\lp=0$.  Surprisingly, this does not mean that first and second sound are uncoupled density and temperature oscillations.  We discuss this in detail in ~\ref{vfieldssec}.

Two-fluid hydrodynamics leads to the coupled equations~\cite{cambridge41,cambridge09}
\beq u^2\delta \rho = \delta P,\;\;u^2\delta\bar{s} = \bar{s}^2_0\frac{\rho_{s0}}{\rho_{n0}}\delta T.\label{LTFeqs}\eeq
As noted above (see figure~\ref{soundfig}), the speeds of first and second sound are well approximated by $c_1$ and $c_2$ in (\ref{cspeeds}).  Physically, this means that to leading order, the pressure fluctuations in first sound are at constant entropy per unit mass $\bar{s}$ and hence, $\delta P \simeq (\partial P/\partial\rho)_{\bar{s}}\delta\rho$.  Similarly, the temperature fluctuations in second sound are to leading order at constant pressure and hence, $\delta T \simeq (\partial T/\partial\bar{s})_{p}\delta\bar{s}$.  Using these in (\ref{LTFeqs}) leads to the leading order expressions for $u_1$ and $u_2$ given in (\ref{uexpansion}).

We remark that, qualitatively, our results in figure~\ref{soundfig} for the temperature dependence of $u_1$ and $u_2$ are in agreement with Arahata and Nikuni~\cite{arahatanik}, who also based their work on NSR thermodynamics.  The differences with \cite{arahatanik} at high and low temperatures are easily understood.  Direct numerical calculations based on NSR are difficult to do accurately for $T < 0.4 T_c$ because of the extremely small value of the normal fluid density. Our results in figure~\ref{soundfig} in this region are based on the assumption that Goldstone phonons are the dominant thermal excitations at low $T$ at unitarity, which allows us to obtain the low temperature values of $u_1$ and $u_2$ analytically. In the opposite limit of high temperatures close to $T_c,$ our calculations show that $W_2$ steadily increases, in contrast with that of \cite{arahatanik}.  For $T$ close to $T_c$, it is crucial to use a superfluid density with the correct critical behaviour~\cite{note0}, removing the spurious first-order behaviour predicted by NSR~\cite{Fukushima07}. (In fact, this behaviour is not unique to NSR and is symptomatic of any theory that treats phase fluctuations in a perturbative way.  See \cite{Taylor09} for further discussion.)

\begin{figure}
\begin{center}
\epsfig{file=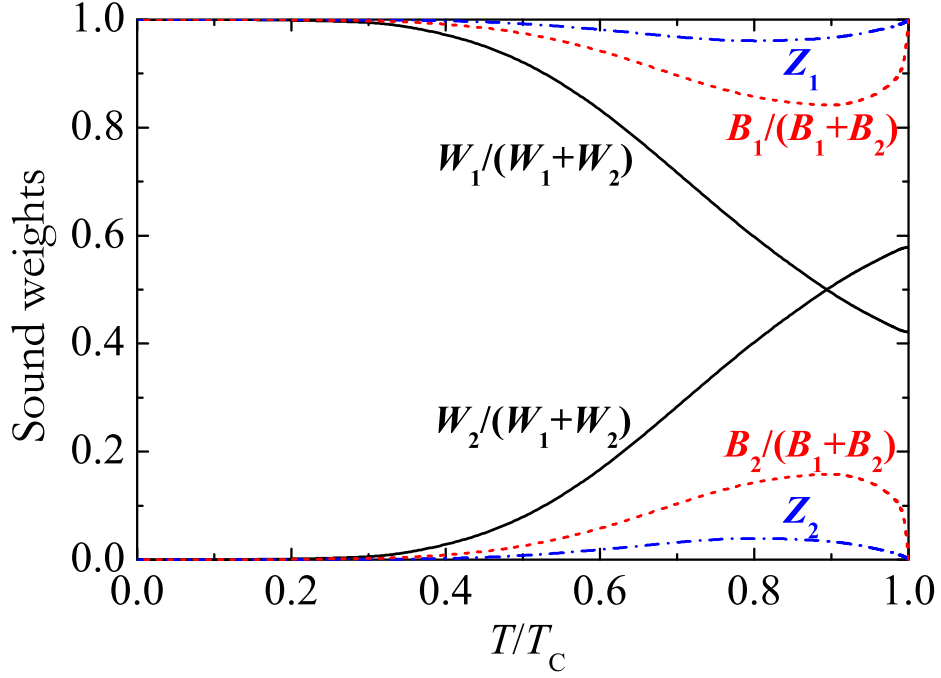, angle=0, width=0.75\textwidth}
\caption{ The temperature dependence of the normalized amplitudes $Z_i$, $B_i$, and $W_i$ of first ($i=1$) and second sound ($i=2$) in a Fermi gas at unitarity.}
\label{Zfig}
\end{center}
\end{figure}

\begin{figure}
\begin{center}
\epsfig{file=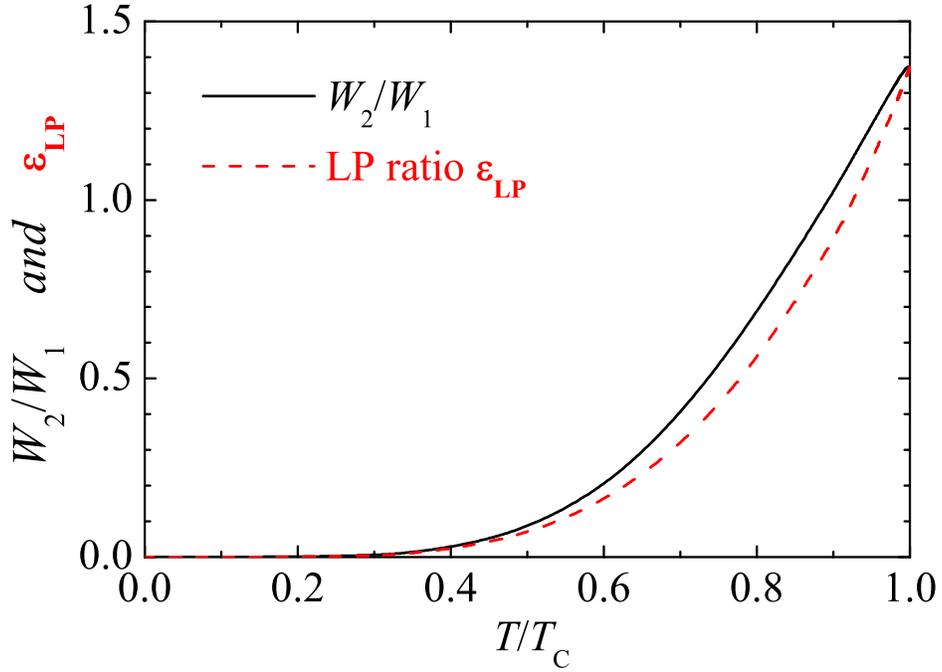, angle=0, width=0.75\textwidth}
\caption{ Comparison between the ratio of the second and first sound amplitudes ($W_2/W_1$) and the Landau-Placzek ratio $\lp=\gamma-1$ in a unitary Fermi gas.}
\label{lpfig}
\end{center}
\end{figure}

Using the values of $u_1, u_2$, and $v$ (see figure~\ref{soundfig}), we can calculate $Z_2, W_2$ and $W_1$. The results are shown in figure~\ref{Zfig}. The relative weight of second and first sound in $S(q,\omega)$ is given by
\beq \frac{W_2}{W_1} = \frac{Z_2 u^2_1}{Z_1 u^2_2} = \frac{v^2-u^2_2}{u^2_1-v^2}\frac{u^2_1}{u^2_2}.\label{W2overW1b}\eeq
Using the exact two-fluid expressions for $u_1$ and $u_2$ gives the ratio $W_2/W_1$ shown in figure~\ref{lpfig}.  For comparison, we also plot the Landau-Placzek ratio $\lp=\gamma-1$, calculated directly from the same thermodynamic functions used to obtain the sound speeds $u_1$, $u_2$, and $v$~\cite{tahugrif08,tayhu09}.

\begin{figure}
\begin{center}
\epsfig{file=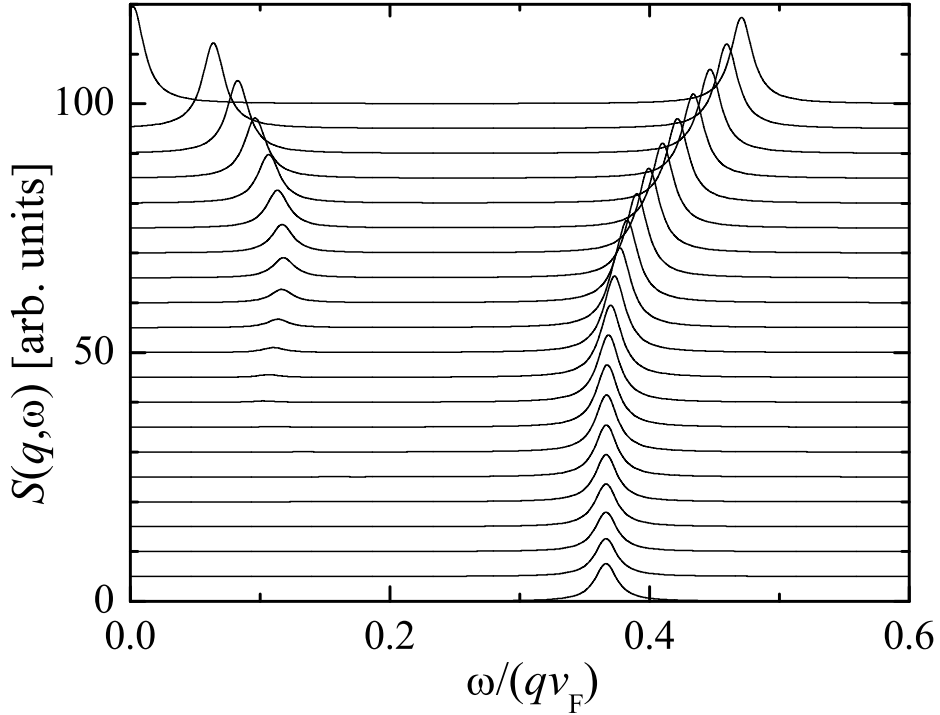, angle=0, width=0.75\textwidth}
\caption{  The two-fluid dynamic structure factor given by (\ref{sqomegahydro}) at unitarity.  The plot shows the first and second sound resonances as a function of the frequency $\omega/qv_F$ (where we take $q=0.1k_F$ and $v_F$ is the Fermi velocity) for a series of temperatures from $T=0$ to $T=T_c$, in steps of $0.05T_c$ (each temperature is offset).  For clarity, the delta functions in (\ref{sqomegahydro}) are plotted as Lorentzians with a width $\Delta = 0.01qv_F$.}
\label{Sqomegafig}
\end{center}
\end{figure}

\begin{figure}
\begin{center}
\epsfig{file=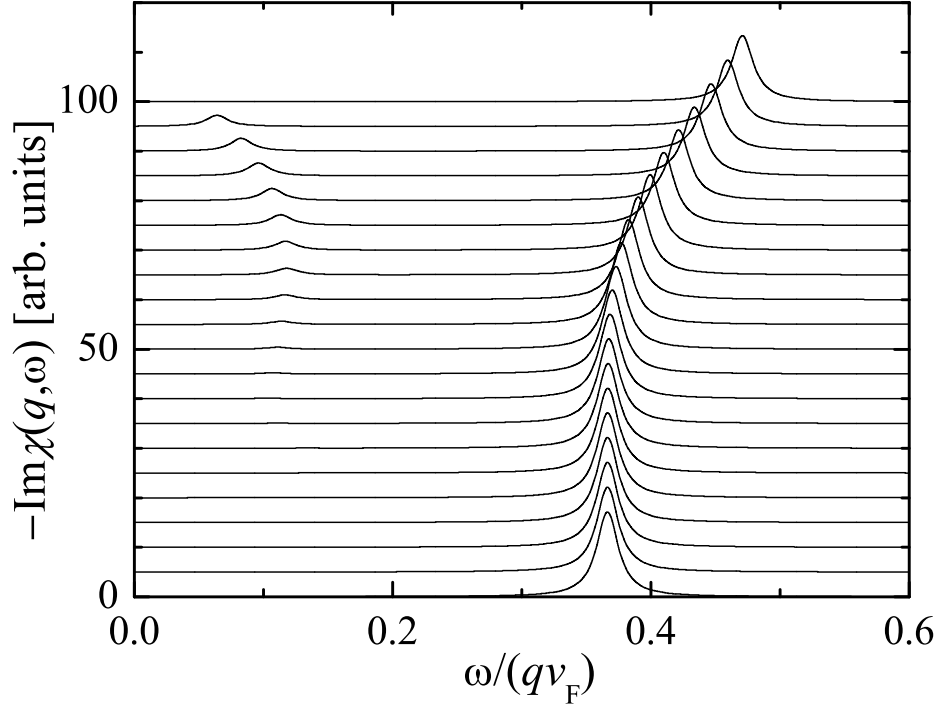, angle=0, width=0.75\textwidth}
\caption{ The imaginary part of the two-fluid density response function at unitarity.  In contrast to the results shown in figure~\ref{Sqomegafig}, where the relative weights of second and first sound are determined by $W_2/W_1$, in Bragg scattering they are determined by the smaller ratio $B_2/B_1$ [see (\ref{Braggweights})].  As a result, second sound has a much smaller weight in Bragg scattering that first sound.}
\label{Chinnfig}
\end{center}
\end{figure}

In figure~\ref{Sqomegafig}, we plot $S(q, \omega)$ given by (\ref{sqomegahydro}).  This shows the first and second sound resonances at a series of temperatures.   The results shown in figures~\ref{Zfig}-\ref{Sqomegafig} show the remarkable feature that, even though the maximum value of $Z_2$ is $\sim 0.05$ at $T\sim0.8 T_c$, the relative weights of first and second sound in $S(q, \omega)$ can be comparable in the temperature region $T\gtrsim 0.8 T_c$.  The relative weight of second sound is instead much smaller in $\mathrm{Im}\chi_{nn}$ (see figure~\ref{Chinnfig}) for reasons we have discussed.  

Figure~\ref{lpfig} also shows that $W_2/W_1$ is quite well approximated by $\lp$.  It is useful to derive this result more analytically.  Using (\ref{speedidentities}) in (\ref{W2overW1b}), we obtain after some algebra
\beq \frac{W_2}{W_1} = \frac{1}{\lp}\left(\frac{u^2_1}{v^2_T}-1\right)^2.\label{W2overW1f}\eeq
This expression was first given by Vinen~\cite{Vinen71}.  Using the expansion for $u^2_1$ in (\ref{uexpansion}), one can reduce (\ref{W2overW1f}) to
\beq \frac{W_2}{W_1} =\lp(1+\gamma x + \cdots)^2.\label{W2overW1c}\eeq
At high temperatures, where $x=c^2_2/c^2_1\ll 1$ (see figure~\ref{soundfig}), we have 
\beq \frac{W_2}{W_1}= \lp[1+2(1+\lp)x+\cdots] \simeq \lp = \gamma-1.\label{W2overW1}\eeq
This result was first obtained almost 40 years ago~\cite{hohmar64,vinen70,Hohenberg73,ferrel68} in the context of Brillouin light scattering~\cite{greytak} in superfluid $^4$He near the critical region close to $T_c$.   
For $T\lesssim 0.4T_c$, one can see from figure~\ref{soundfig} that we can no longer assume $x\ll 1$ and hence expand (\ref{W2overW1c}). 

As noted above, the result in (\ref{W2overW1}) was obtained in the classic literature on superfluid $^4$He. The Landau-Placzek ratio $\lp$ is extremely small $(10^{-2}-10^{-3})$ in superfluid $^4$He and thus second sound has generally a very small amplitude in the dynamic structure factor. However, under pressure and close to $T_c$, the magnitude of $\lp$ in liquid $^4$He can be significantly increased (to about $\lp\sim0.2$) such that a second sound doublet in $S(q, \omega)$ can be measured using Brillouin light scattering techniques~\cite{greytak}. Such studies~\cite{winterling73,connor75} have verified the correctness of $W_2/W_1=\lp$ in liquid $^4$He at a pressure of $P=25$ atmospheres in the temperature region $(T_c-T)\sim 10^{-2}-10^{-4}K.$

The Landau-Placzek ratio $\lp\equiv \gamma-1$ was first derived in 1934~\cite{lanplac34} to describe the relative weight of the thermal diffusion central peak vs. sound waves which are exhibited by $S(q, \omega)$ describing the hydrodynamics of a normal fluid. Thus, (\ref{W2overW1}) generalizes the Landau-Placzek result~\cite{lanplac34,mazenko06} to the case of two-fluid hydrodynamics.
In a normal liquid above $T_c$, the dynamic structure factor given by (\ref{sqomegahydro}) reduces to~\cite{mazenko06}
\beq S(q,\omega) = \frac{T}{2m}\Bigg\{\frac{1}{v^2_{\bar{s}}}\left[\delta(\omega-v_{\bar{s}}q)+\delta(\omega+v_{\bar{s}}q)\right]+ \frac{\gamma-1}{v^2_{\bar{s}}}\delta(\omega)\Bigg\}.\label{sqomeganormal}\eeq
Thus, above $T_c$, the second sound doublet in (\ref{sqomegahydro}) collapses into a zero frequency central peak, with relative weight $\gamma-1$ compared to a sound wave at $\omega = v_{\bar{s}}q$.  When one includes damping due to transport processes, the central peak in (\ref{sqomeganormal}) broadens, describing the thermal diffusion relaxation mode of width $D_Tq^2$ (where $D_T$ is the thermal diffusivity).  The behaviour of $S(q,\omega)$ as one goes from below $T_c$ to above $T_c$ is smooth but complicated due to various singularities in the thermodynamic and transport coefficients (for further discussion, see \cite{vinen70}.) 

\section{Dynamic structure factor of dilute Bose gases}
\label{Bosegassec}

In this section, we discuss first and second sound in a uniform dilute
Bose-condensed gas. The two-fluid hydrodynamics is again described by the
results given in section~\ref{sqomegatwofluidsec}. However, we need to distinguish between the
cases of weakly and strongly interacting Bose gases. The latter
case can be easily realized by considering a molecular Bose condensate on
the BEC side of the BCS-BEC crossover in a two-component Fermi gas. In such
a molecular Bose gas with density $n_B=n/2$ and molecular mass $m_B=2m$, the \textit{s}-wave
molecular scattering length is given by $a_B\simeq 0.6a_F$, where $a_F$ is the atomic $s$-wave scattering length between the two Fermi components~\cite{PetrovMol}. We will show that the behaviour of first and second
sound in strongly interacting Bose gases is very similar to that in Fermi gas
superfluids near unitarity, but very different from that in weakly interacting Bose gases.

In a dilute, weakly interacting Bose gas, the thermodynamic quantities that enter the two-fluid
equations can be solved analytically in powers of the interaction strength $%
g\equiv 4\pi a_B/m_B$. To first order in $g$, the speeds of first and second
sound are given by (see, for instance, chapter~15 in \cite{cambridge09}) 
\beq
u_1^2 =\frac{5T}{3m_B}\frac{g_{5/2}\left( 1\right) }{g_{3/2}\left(
1\right) }+\frac{2g\tilde{n}_0}{m_B},\;\;\;
u_2^2 =\frac{gn_{c0}}{m_B}. \label{uexpansionBEC}
\eeq
Here, $g_n(z)=\sum_{l=1}^\infty z^l/l^n$ is the usual Bose-Einstein function
with fugacity $z$. In the limit of small $g$, the normal fluid density $\rho
_{n0}$ reduces to the density $m\tilde{n}_0\equiv \rho -mn_{c0}$ of atoms
thermally excited out of the condensate. The superfluid density $\rho _{s0}$
is likewise given by the density $mn_{c0}$ of condensate atoms.  To lowest order in $g$, $n_{c0}$ is the condensate density of the ideal Bose gas.

\begin{figure}
\begin{center}
\epsfig{file=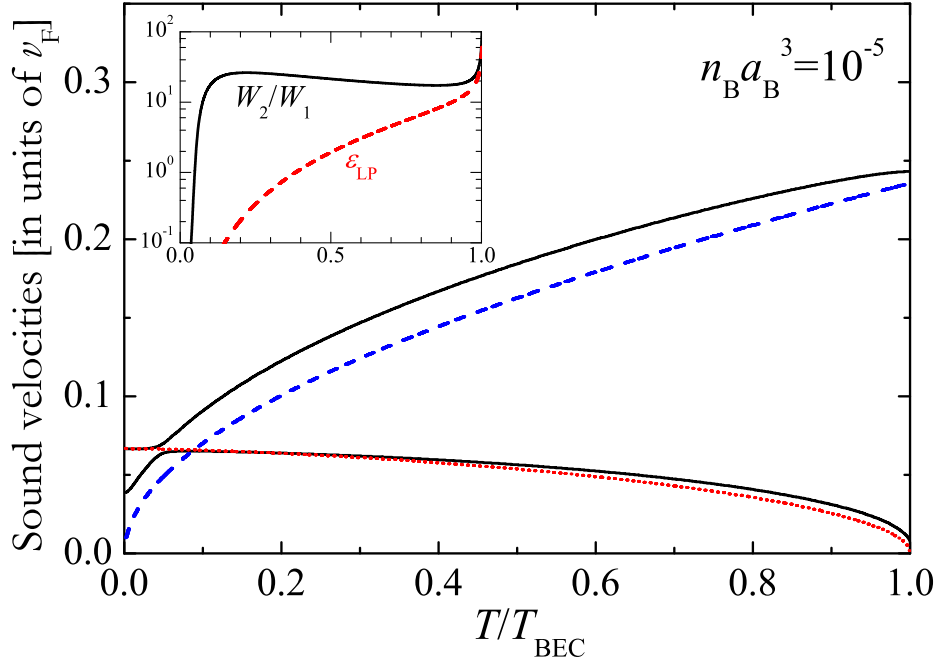, angle=0, width=0.75\textwidth}
\caption{ Speeds of first and second sound in a weakly
interacting molecular Bose gas with gas parameter $n_Ba_B^3=10^{-5}$. The blue
dashed and red dotted lines are respectively the approximate speeds $u_1$
and $u_2$ given by (\ref{uexpansionBEC}). The inset shows the LP ratio  $\lp$
(red dashed line) and the weight ratio $W_2/W_1$ (black solid line).}
\label{fig6} 
\end{center}
\end{figure}

In figure~\ref{fig6}, we show the temperature dependence of $u_1$ and $u_2$, satisfying (\ref{speedidentities}), in the case of weak interactions ($n_Ba_B^3=10^{-5}$). The thermodynamic
functions used to determine these sound velocities have been calculated using the Hartree--Fock--Popov
(HFP) microscopic model for the thermal excitations~\cite{cambridge09}.  The velocities are normalized in terms of a Fermi
velocity defined by $v_F=(6\pi ^2n_B)^{1/3}/m$ (equivalent to the Fermi velocity of a two component Fermi gas through the BCS-BEC crossover).   We also show for comparison the
values of $u_1$ and $u_2 $ given by the lowest order solutions described by (\ref{uexpansionBEC}). One sees that these uncoupled modes of oscillations are a good
first approximation to the full solutions of the two-fluid equations in a
weakly interacting BEC at all temperatures outside a small temperature
region at low temperatures where first and second sound hybridize,
exhibiting an avoided crossing. At this avoided crossing, the natures of
first and second sound are interchanged. As $T\rightarrow 0$, the first
sound velocity coincides with the phonon velocity, as is the case with
strongly interacting Fermi gases and superfluid $^4$He. For a dilute gas,
this phonon velocity is the Bogoliubov--Popov phonon velocity $\sqrt{gn_{c0}/m_B}$, and corresponds to an oscillation of the condensate
component. Above the hybridization point, to a good approximation, second
sound propagates with this velocity. Consequently, second sound in a dilute
Bose gas (above the hybridization temperature) is essentially an oscillation
of the condensate at all temperatures, with a static thermal cloud. In
contrast, first sound describes an oscillation of the thermal cloud in the presence of a
stationary condensate.

%


We note that HFP approximation we have used to calculate thermodynamic quantities has the well-known problem near $T_c$ of predicting a spurious first-order transition (see, for instance,  \cite{ShiGriffin}).  As touched on in section~\ref{sqomegaweightssec}, this is actually a very general problem, and arises in any theory that treats bosonic phase fluctuations of the order parameter as a small parameter, including the HFP and NSR theories.  Of course, this includes all perturbation theories that go beyond mean-field and only \textit{ab initio} Quantum Monte-Carlo calculations can avoid this problem.  (A competing fluctuation theory of the BCS-BEC crossover~\cite{He07} avoids this problem by ignoring phase fluctuations.) The unphysical first-order transition leads to the result that the second sound velocity does not vanish at $T_c$ as it should.  In contrast to figure~\ref{soundfig}, we have not corrected the spurious behaviour of the superfluid (condensate) density close to $T_c$ in the results shown in figures~\ref{fig6} and {\ref{fig7}. These figures also show the expected feature that the error becomes larger with increasing interaction strength. In spite of the shortcomings of HFP theory in the critical region close to $T_c$, we emphasize that it gives a good estimate of thermodynamic quantities, outside the critical region.  All numerical results shown in figures~(\ref{fig6})-(\ref{fig8}) are obtained using this theory (apart from the uncoupled solutions shown in figure~\ref{fig6}, which use the ideal gas expression for the condensate density).  

In figure~\ref{fig6}, the inset shows the much larger value of the LP ratio $\lp$ (compared to a Fermi superfluid--see figure~\ref{lpfig}) and the ratio $W_2/W_1$ of the amplitudes of second and first sound in $S(q,\omega) 
$ as a function of the temperature. We note that ratio $W_2/W_1$ is still
given by the expressions in (\ref{W2overW1b}) and (\ref{W2overW1f}). However, now the parameter $\lp$ is no longer small ($\ll 1$) and thus the expansion based
on (\ref{uexpansion}), which leads to (\ref{W2overW1c}) and (\ref{W2overW1}), is no longer valid.  We also explicitly note that $W_2/W_1$ is not equal to $\lp$ in either the weak or strong coupling limits.

First and second sound in a dilute, weakly interacting Bose gas involve weakly coupled oscillations of the thermal cloud and condensate, respectively. Both modes contribute to the density response function.  Consequently, both first and second sound are sensitive to density probes. This
fact has been made use of in a recent experiment involving trapped atomic Bose
gases by Meppelink {\it et al}.~\cite{meppelink09}. In this study, a density pulse
was generated (as described in section~\ref{sqomegasec}) and the velocity of the second sound pulse was measured above
the hybridization point by tracking the propagation of the density pulse.  
The inset in figure~\ref{fig6} shows that the amplitude of second sound in $S(q,\omega)$ and in density pulse propagation experiments is twenty times as strong as first sound for all temperatures above the crossing point
at $T\sim 0.1T_c$.  This result may explain the absence of a first sound pulse in the data of \cite{meppelink09}, which corresponded to $n_Ba_B^3 \sim 5\times 10^{-6}$.   

%
\begin{figure}
\begin{center}
\epsfig{file=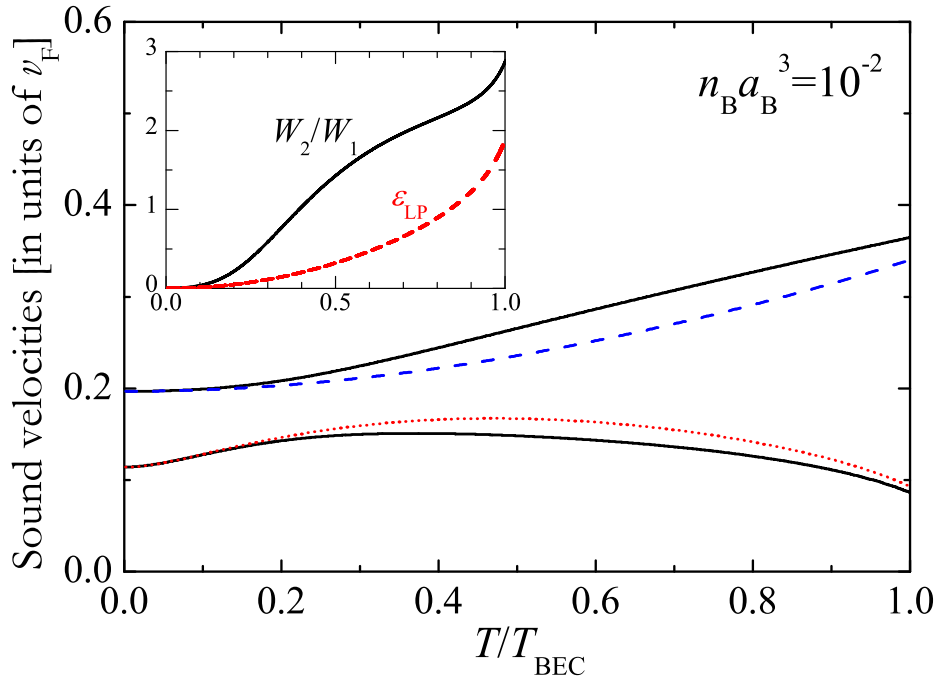, angle=0, width=0.75\textwidth}
\caption{ Speeds of first and second sound in a strongly
interacting molecular Bose gas with gas parameter $n_Ba_B^3=10^{-2}$. The blue
dashed and red dotted lines are respectively the approximate speeds $%
c_1=\upsilon _{\bar{s}}$ and $c_2=\upsilon /\sqrt{\gamma }$ given by (\ref{cspeeds}). The inset shows the LP ratio $\lp$ (dashed line) and the
weight ratio $W_2/W_1$ (black line), to be compared with the result in figure~\ref{lpfig} for a strongly interacting Fermi gas at unitarity.}
\label{fig7} 
\end{center}
\end{figure}


In figure~\ref{fig7}, we plot first and second sound for a more strongly interacting
Bose gas (i.e., with a larger value of $g$.) This situation is close to the
unitarity Fermi gas results shown in figure~\ref{soundfig}~\cite{ArahataComment}. In particular, we see that the
velocities of first and second sound never become degenerate (cross) and
thus the hybridization shown in figure~\ref{fig6} does not occur. In this
strongly interacting Bose superfluid, we also show the approximate
velocities of $u_1$ and $u_2$ given by (\ref{cspeeds}). These approximate values
give a reasonable first order approximation to the full solutions of the
first and second sound velocities, rather than the expression in (\ref{uexpansionBEC})
that applies to a weakly interacting Bose gas as shown in figure~\ref{fig6}. The inset
shows the ratio of $W_2/W_1$ vs. $T$, as computed from (\ref{W2overW1b}). In this case, the amplitude of second sound in $S\left(q,\omega \right) $ is only 2-3 times larger than first sound for $T\gtrsim 0.5T_c$.

In figure~\ref{fig8}, we compare the values of $W_1/(W_1+W_2)$ and $W_2/(W_1+W_2)$ for a
weakly interacting Bose gas ($n_Ba_B^3=10^{-4}$) and a strongly interacting
Bose gas ($n_Ba_B^3=0.1$). One sees that the latter case is qualitatively
similar to the result in a Fermi superfluid gas at unitarity given in figure~\ref{Zfig}. In contrast, in a weakly interacting Bose gas, for the temperature range
above the hybridization point, the amplitude of second sound is much larger
than first sound in both $S(q,\omega)$ and density pulse
propagation experiments.

%
\begin{figure}
\begin{center}
\epsfig{file=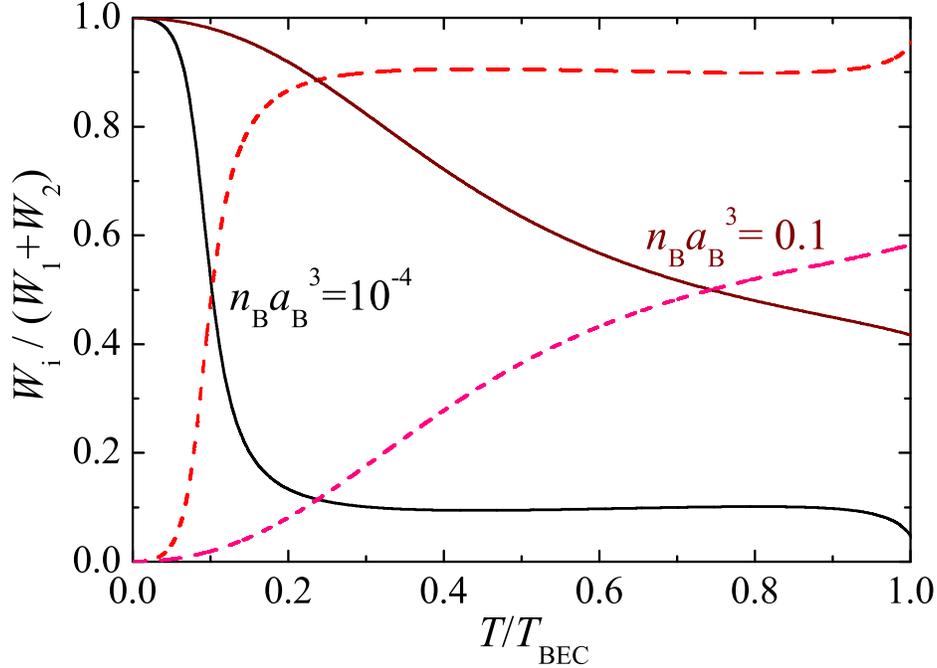, angle=0, width=0.75\textwidth}
\caption{Temperature dependence of the normalized amplitudes $%
W_1/(W_1+W_2)$ (solid lines) and $W_2/(W_1+W_2)$ (dashed lines) for a
weakly interacting Bose gas ($n_Ba_B^3=10^{-4}$) and a strongly interacting
Bose gas ($n_Ba_B^3=0.1$).}
\label{fig8} 
\end{center}
\end{figure}


\section{Bragg scattering in a  trapped atomic gas}
\label{Braggsec}

For illustration, we now consider the response of a trapped Fermi gas at unitarity to two-photon Bragg scattering due to first and second sound. 
We consider an experimental setup similar to that used in \cite{Miller07}, where a pair of laser beams are applied to the gas such that the region where they overlap is confined to the centre of the gas.  As a first approximation to this configuration, we assume that the overlap region between the two beams is isotropic, with a Gaussian profile:
\beq \delta V(\bfr,t) = \frac{V}{\sqrt{2\pi} \sigma}e^{-r^2/2\sigma^2}\cos(\bq\cdot\bfr-\omega t).\eeq
Here, $\sigma$ is the width of the Bragg beams and $\bq$ and $\omega$ are the wavevector and frequency difference between the two beams.

\begin{figure}
\begin{center}
\epsfig{file=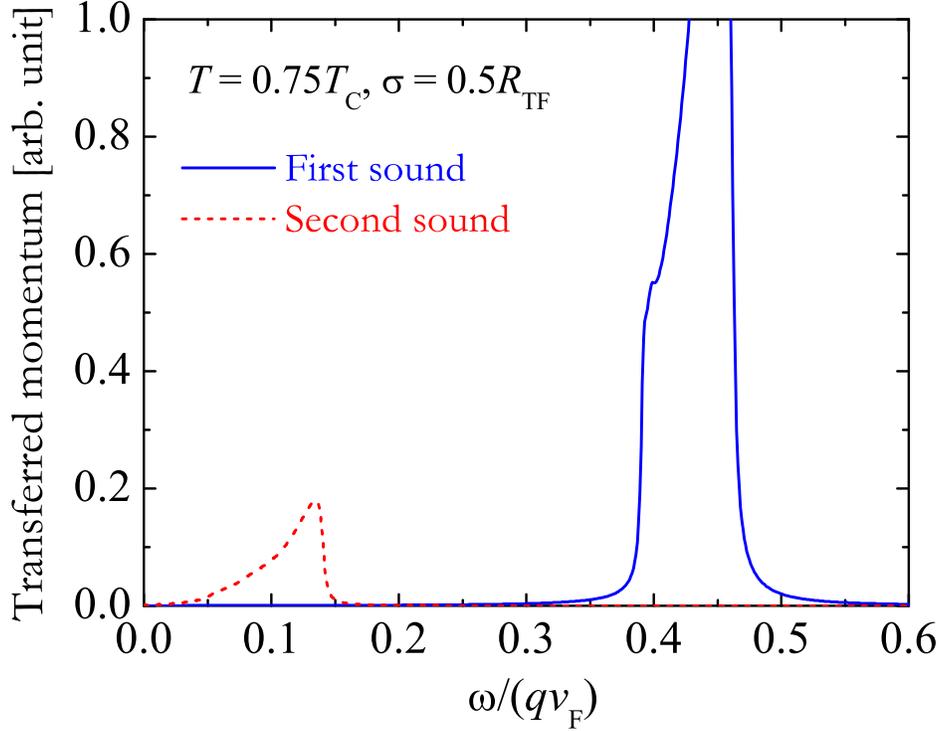, angle=0, width=0.75\textwidth}
\caption{Momentum transferred to a trapped Fermi gas at unitarity by two photon Bragg scattering due to first and second sound at $T\simeq 0.75T_c$.  Solid (blue) line shows the momentum transferred due to first sound as a function of the beam detuning $\omega$ divided by the relative wave vector $q\equiv |\bq_1-\bq_2|$.  $v_F$ is the Fermi velocity.  Dashed (red) line shows the momentum transferred due to second sound. }
\label{Braggfig}
\end{center}
\end{figure}
  
For a Bragg pulse of short duration ($\tau \ll 2\pi/\omega_0$, where $\omega_0$ is the trap frequency), the momentum transferred to the gas is related to the imaginary part of the density response function [see (\ref{DeltaPuniform})], through a convolution of the atomic density:
\beq \Delta P(q,\omega) =  \frac{q V^2 \tau}{4\pi N\sigma^2}\int d^3r\; n(r)e^{-r^2/\sigma^2}\mathrm{Im}\chi_{nn}(q,\omega;r)
\label{deltaP}\eeq
In this expression, $\mathrm{Im}\chi_{nn}(q,\omega;r)$ is the imaginary part of the density response function, evaluated within a local density approximation (LDA). 
The imaginary part of the two-fluid density response function is given by (\ref{Imchinntwofluid}) and should be evaluated at the local value of the density.  The LDA  is justified for wave vectors much larger than the inverse of the size of the trapped gas, so that the system can be considered, locally, as a uniform body.

Results for the momentum transferred given by (\ref{deltaP}), as a function of $\omega/qv_F$, are shown in figure~\ref{Braggfig} for $T=0.75T_c$ and choosing $\sigma =0.5R_{TF}$, where $R_{TF} \equiv \sqrt{\hbar/m\omega}(24 N)^{1/6}$ is the Thomas--Fermi radius of a noninteracting two-component Fermi gas.  Although small, the second sound peak is well separated from the first sound peak, revealing that two-photon Bragg spectroscopy might become a practical way to measure second sound.  Above and below this temperature, the contribution of second sound to the momentum transfer becomes significantly  smaller, however.

\section{Conclusions}

In this paper, we have given a detailed analysis of the density response function $\chi_{nn}(q,\omega)$ and the related dynamic structure factor $S(q,\omega)$ for uniform superfluids in the two-fluid hydrodynamic limit.  We have put special emphasis on the second sound contribution to $S(q,\omega)$ in the case of a strongly interacting Fermi gas superfluid (at unitarity).  While many features of the dynamic structure factor we discuss are known in the superfluid $^4$He literature~\cite{greytak,vinen70}, several new aspects arise in the case of ultracold superfluid gases.  

First and second sound both appear as resonances in the density response function but their relative weights vary in different experimental probes.  Two-photon Bragg scattering, a well-developed tool in ultracold gases used to study density fluctuations (so far, mainly in the collisionless region~\cite{StamperKurn99,Katz04,Davidson05}), is directly proportional to $\mathrm{Im}\chi_{nn}(q,\omega)$.  In contrast, the low-frequency limit of the dynamic structure factor $S(q,\omega) = -\mathrm{Im}\chi_{nn}(q,\omega)/\pi n[1-\exp(-\beta\omega)]$ measured in Brillouin light scattering, and also the amplitude of density pulses that can be generated in dilute gases~\cite{arahatanik}, are proportional to $\mathrm{Im}\chi_{nn}(q,\omega)/\omega$.  This extra factor of $1/\omega$ leads to a large enhancement of the weight of low frequency second sound in $S(q,\omega)$ compared to first sound.

The relative spectral weights of first and second sound in $\chi_{nn}(q,\omega)$ are very dependent on how strongly temperature fluctuations couple to density fluctuations.  For $S(q,\omega)$, this coupling is conveniently parametrized by the value of the dimensionless Landau--Placzek ratio $\lp = \gamma-1$, where $\gamma$, defined in (\ref{gammadef}), is the ratio of the specific heats (per units mass) at constant pressure and volume.  At finite temperatures, the relative weight of first and second sound ($W_2/W_1$) is equal to $\lp$ in both unitary Fermi gases and superfluid $^4$He.  The key difference between superfluid $^4$He and a unitary Fermi superfluid lies in the value of $\lp$.  Apart from a small region very close to the transition temperature $T_c$, one has $\lp \ll 1$ in superfluid $^4$He~\cite{winterling73,connor75}.  In contrast, for $T\gtrsim 0.7T_c$, one finds that $\lp$ is of order unity in Fermi superfluids (see figures~\ref{lpfig} and \ref{Sqomegafig}) and hence the second sound resonance has appreciable weight in $S(q,\omega)$.

Density pulse experiments as suggested in \cite{arahatanik} seem to be a promising way of detecting both first and second sound pulses in Fermi gases, although more work is needed to clarify the nature of first and second sound in cylindrical geometries where such pulses would be generated. (See, for instance, recent work on this problem in \cite{Bertaina10}.)  The experimental results reported in ~\cite{Joseph07}, where a blue-detuned laser was used to generate a density pulse in a Fermi gas close to unitarity, revealed no clear signs of a second sound pulse.  This experiment was carried out at very low temperatures, however, where second sound will not have appreciable weight in $S(q,\omega)$ (see figures~\ref{lpfig} and \ref{Sqomegafig}).  Alternately, even though the weight of second sound is comparatively small in Bragg scattering, Bragg scattering has the advantage that it can be used to probe in a very precise way the properties of uniform superfluid Fermi gases~\cite{Miller07}, including the velocity of second sound.  

In section~\ref{Bosegassec}, we compare the two-fluid hydrodynamics in a uniform, dilute, weakly interacting Bose condensed gas (figure~\ref{fig6}) with a strongly interacting Fermi superfluid gas.  In such Bose superfluids, the thermal expansion (and hence, $\lp$) is much larger, with the result that second sound is the dominant excitation in $S(q,\omega)$ and mainly involves a pure oscillation of the condensate in the presence of a static thermal component~\cite{cambridge09,meppelink09}.  However, first and second sound in a strongly interacting Bose gas are similar to those in a Fermi gas near unitarity (see figure~\ref{fig7}).  


\begin{acknowledgments}
We thank Lev Pitaevskii for useful comments.  E.T. was supported by NSF-DMR 0907366 and ARO W911NF-08-1-0338.  H.H. and X.-J.L. were supported by ARC and NSFC.  S.S. acknowledges the support of the EuroQUAM FerMix program.  A.G. was supported by NSERC and CIFAR.
\end{acknowledgments}

\appendix

\section{Phonon thermodynamics at low temperatures}
\label{phononthermosec}

At low temperatures, Goldstone phonons determine the thermodynamics of both superfluid $^4$He and Fermi gases.  For phonons with velocity $c$, the free energy $F$ and normal fluid density $\rho_{n0}$ are given by~\cite{LLSM2} (Recall that we have set $\hbar=k_B=1$)
\beq F = F_0 - \frac{V\pi^2T^4}{90c^3}\label{F}\eeq
and
\beq \rho_{n0} = \frac{2\pi^2 T^4}{45 c^5}.\label{rhon}\eeq
Using (\ref{F}), it is straightforward to show that 
\beq \bar{s}_0 = -\frac{1}{mN}\left(\frac{\partial F}{\partial T}\right)_{V,N}=\frac{2\pi^2T^3}{45\rho c^3},\label{sbar}\eeq
\beq P \!=\! -\left(\frac{\partial F}{\partial V}\right)_{T,N}\!\!\! = \!P_0 \!+\! \frac{\pi^2T^4}{90c^3}\left[1+\frac{3\rho}{c}\left(\frac{\partial c}{\partial\rho}\right)_{\!T,N}\right],\label{P}\eeq
and
\beq \bar{c}_{v} = \frac{2\pi^2 T^3}{15\rho c^3}.\label{cv}\eeq
In arriving at these expressions, the temperature dependence of the Goldstone phonon velocity $c$ has been ignored.  $P_0$ is the pressure in the ground state.  
These results can be combined to give
\beq \bar{c}_{p}=\bar{c}_{v} - T\left(\frac{\partial{\bar{s}}}{\partial\rho}\right)_{\!T}\left(\frac{\partial P}{\partial T}\right)_{\!\rho}\left(\frac{\partial P}{\partial\rho}\right)^{-1}_{\!T}=\bar{c}_{v} + \frac{4\pi^4 T^7}{45^2\rho^2c^6v^2_T}\left[1 + \frac{3\rho}{c}\left(\frac{\partial c}{\partial\rho}\right)_{\!T}\right]^2,\label{cp}\eeq
and
\beq v^2\equiv\bar{s}^2_0\frac{T}{\bar{c}_v}\frac{\rho_{s0}}{\rho_{n0}}=\frac{c^2}{3}\frac{\rho_{s0}}{\rho_{0}}.\label{v}\eeq

Combining the above results, the Landau--Placzek ratio  at low temperatures is given by
\beq \lp = \frac{2\pi^2T^4}{135\rho c^3 v^2_T}\left[1+\frac{3\rho}{c}\left(\frac{\partial c}{\partial\rho}\right)_{\!T}\right]^2\label{lp}\eeq
and hence, recalling that $x = v^2/\gamma v^2_{\bar{s}}$ and $\gamma = v^2_{\bar{s}}/v^2_T$, 
\beq \sqrt{\frac{\rho_{s0}}{\rho_{n0}}\lp x} = \frac{1}{3} \frac{\rho_{s0}}{\rho_{0}}\frac{c^2}{v^2_{\bar{s}}}\left[1 + \frac{3\rho}{c}\left(\frac{\partial c}{\partial\rho}\right)_{\!T}\right].\label{smallparameteridentity1}\eeq
This result is valid for both the superfluid unitary Fermi gas and superfluid $^4$He.  Further simplifications are not possible without \textit{ab-initio} or experimental information about $c$.  At low temperatures, however, where $\rho_{s0}\simeq \rho_0$ and $v_{\bar{s}}\simeq c$ (see figure~\ref{soundfig}),  (\ref{smallparameteridentity1}) reduces to
\beq \sqrt{\frac{\rho_{s0}}{\rho_{n0}}\lp x} \simeq \frac{1}{3}\left[1 + \frac{3\rho}{c}\left(\frac{\partial c}{\partial\rho}\right)_{\!T}\right].\label{smallparameteridentity2}\eeq
Equations~(\ref{cp}) and (\ref{v}) also give us the result that $\gamma \to 1$ and $x = v^2/\gamma v^2_{\bar{s}}\to 1/3$ as $T\to 0$.  
Finally, for a Fermi gas at unitarity, $(\rho/c)(\partial c/\partial\rho)_T = 1/3$ and (\ref{smallparameteridentity2}) is further reduced to the result given in (\ref{smallparameteridentity4}).   In superfluid $^4$He, the Gr\"uneisen constant is almost a factor of 9 larger, $(\rho/c)(\partial c/\partial\rho)_T\simeq 2.84$~\cite{Abraham70}.

\section{Superfluid and normal fluid velocity fields in second sound}
\label{vfieldssec}

As Landau~\cite{cambridge41} originally noted, when the thermal expansion coefficient can be neglected (i.e., $\lp=0$), one can show that second sound is a pure temperature oscillation, with no associated density fluctuations.  This last feature follows from the fact that when $\lp=0$, second sound involves no mass current ($\bj \equiv \rho_{n0}\bv_n + \rho_{s0}\bv_s=0$) and thus the continuity equation requires that $\delta\rho=0$.  Likewise, first sound is a pure density oscillation, and involves no fluctuations in the entropy per unit mass $\bar{s}$.  The non-dissipative two-fluid equations give the following equation for this quantity:
\beq \frac{\partial\delta\bar{s}}{\partial t} = \frac{\bar{s}_0\rho_{s0}}{\rho_0}\bnab\cdot(\bv_s-\bv_n).\label{sbarfluct}\eeq
This shows that for pure adiabatic motion ($\delta\bar{s}=\delta T=0$) such as first sound, the superfluid and normal fluid velocities are identical: $\bv_s = \bv_n$. 
Putting these results together, the velocity fields for first and second sound when $\lp=0$ are given by
\beq \frac{v^{(1)}_s}{v^{(1)}_n}=1,\;\; \frac{\rho_{s0}v^{(2)}_s}{\rho_{n0}v^{(2)}_n} = -1,\label{vfields0}\eeq
where, for instance, $v^{(1)}_{s}$ denotes the superfluid velocity field associated with first sound (along the direction of propagation $\bq$).  This Appendix discusses how these velocity ratios change when $\lp$ is non-zero.  Surprisingly, we find that changes are quite substantial at all temperatures, including the low-$T$ region where $\lp \ll 1$.  

As we discuss in the present paper and in \cite{tayhu09}, the Landau--Placzek ratio $\lp$ increases with temperature and in Fermi gases near unitarity, $\lp$ is of order unity at high temperatures (see figure~\ref{lpfig}).  As we have shown, this is large enough that second sound has significant weight in the dynamic structure factor.  

To understand the effect that a non-zero value of $\lp$ has on the superfluid $\bv_s$ and normal fluid $\bv_n$ velocity fields, one can solve the two-fluid equations for the values of these fields associated with  first and second sound $\omega = uq$, where $u=u_1$ or $u_2$.  Defining the variables $\bj = \rho_{s0}\bv_s + \rho_{n0}\bv_n$ and $\bw = \rho_{s0}(\bv_s-\bv_n)$, the plane-wave solutions of the linearized two-fluid equations are
\bea  \left(\begin{array}{cc}u^2- v^2&-s_0\frac{\rho_{s0}}{\rho_{n0}}\left.\frac{\partial T}{\partial \rho}\right|_{\bar{s}}
\\
s_0\left.\frac{\partial T}{\partial \rho}\right|_{\bar{s}}
&u^2-v^2_{\bar{s}}
\end{array} \right )
 \left(
\begin{array}{c}\bw
\\
\bj
\end{array}\right) = 0.\label{tfsolns}\eea
Here,
\beq s_0\left.\frac{\partial T}{\partial\rho}\right|_{\bar{s}}= \frac{\bar{s}_0}{\rho_0}\left.\frac{\partial P}{\partial\bar{s}}\right|_{\rho} = 
\sqrt{\frac{\rho_{n0}}{\rho_{s0}}} \sqrt{\lp x} v^2_{\bar{s}} \equiv b^2\label{b}\eeq
has units of velocity squared. Recall that $x$ is defined below (\ref{uexpansion}).  Standard thermodynamic identities have been used to introduce $\lp=\gamma-1$.  As the structure of (\ref{tfsolns}) makes clear, $b^2$ determines the coupling between density ($\propto \bj$) and entropy per unit mass ($\propto \bw$) fluctuations.  [The latter is given by (\ref{sbarfluct}).]

Using (\ref{tfsolns}), it is straightforward to derive an expression for the ratio of the velocity fields associated with first and second sound:
\beq \frac{v_s}{v_n} = \frac{s_0\left.\frac{\partial T}{\partial\rho}\right|_{\bar{s}} - \frac{\rho_{n0}}{\rho_{s0}}(u^2-v^2_{\bar{s}}) }{s_0\left.\frac{\partial T}{\partial\rho}\right|_{\bar{s}}+(u^2-v^2_{\bar{s}})}.\label{vsovervn1}\eeq
Substituting the expressions for the speeds of first and second sound given by (\ref{uexpansion}) into (\ref{vsovervn1}), one finds
\beq   \frac{v^{(1)}_s}{v^{(1)}_n} \simeq \frac{v^2_{\bar{s}} - b^2}{v^2_{\bar{s}}+ \frac{\rho_{s0}}{\rho_{n0}}b^2}\label{vsovervn2}\eeq
and
\beq \frac{\rho_{s0}v^{(2)}_s}{\rho_{n0}v^{(2)}_n} \simeq -\frac{v^2_{\bar{s}}(1-x) + \frac{\rho_{s0}}{\rho_{n0}}b^2}{v^2_{\bar{s}}(1-x) - b^2}.\label{vsovervn3}\eeq
In section~\ref{sqomegaweightssec}, we showed that $c_1$ and $c_2$ in (\ref{cspeeds}), which are expansions in the small parameter $\lp x$, are excellent approximations to the speeds of first and second sound at \textit{all} temperatures (see figure~\ref{soundfig}).  We thus expect (\ref{vsovervn2}) and (\ref{vsovervn3}) to be similarly good approximations for the velocity fields at all temperatures.  Using (\ref{b}) to remove the dependence of these expressions on $v_{\bar{s}}$, they reduce to
\beq   \frac{v^{(1)}_s}{v^{(1)}_n} \simeq \frac{1 - \sqrt{\frac{\rho_{n0}}{\rho_{s0}}\lp x}}{1+ \sqrt{\frac{\rho_{s0}}{\rho_{n0}}\lp x}} \label{vsovervn4}\eeq
and
\beq \frac{\rho_{s0}v^{(2)}_s}{\rho_{n0}v^{(2)}_n} \simeq -\frac{(1-x) + \sqrt{\frac{\rho_{s0}}{\rho_{n0}}\lp x}}{(1-x) - \sqrt{\frac{\rho_{n0}}{\rho_{s0}}\lp x}}.\label{vsovervn5}\eeq
When $\lp$ is set to zero, we recover the results in (\ref{vfields0}), describing pure in-phase and out-of-phase oscillations of the superfluid and normal fluid components.  However, we note that even though $\lp$ vanishes as $T\to 0$, so does the normal fluid density $\rho_{n0}$, and it is not obvious that $\sqrt{\rho_{s0}\lp x/\rho_{n0}}$ is small at low temperatures.  Conversely, at high temperatures, $T\to T_c$, the factor of $\sqrt{\rho_{n0}/\rho_{s0}}$ multiplying $\sqrt{\lp x}$ in (\ref{vsovervn4}) and (\ref{vsovervn5}) will enhance the small parameter $\sqrt{\lp x}$.   It is thus crucial to know the temperature dependencies of $\rho_{s0}/\rho_{n0}$ and $\lp x$ in both limits.

\begin{figure}
\begin{center}
\epsfig{file=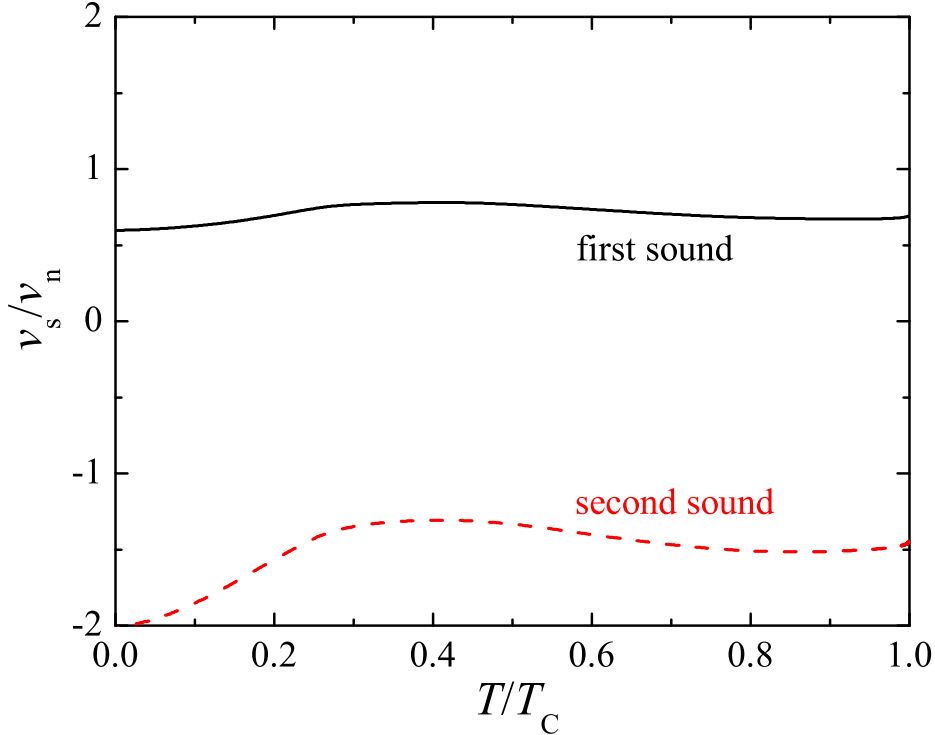, angle=0,width=0.75\textwidth}
\caption{Superfluid and normal fluid velocity fields associated with first and second sound in a unitary Fermi superfluid.  The solid (black) line shows the ratio $v^{(1)}_s/v^{(2)}_n$ of the superfluid and normal fluid velocity fields involved with first sound, (\ref{vsovervn4}).  The dashed (red) line shows the ratio $\rho_{s0}v^{(2)}_s/\rho_{n0}v^{(2)}_n$ of the superfluid and normal fluid currents associated with second sound, (\ref{vsovervn5}).  If we set $\lp=0$, these ratios would be $1$ and $-1$, respectively, describing the velocity fields for pure density and temperature oscillations.}
\label{velocityratiofig}
\end{center}
\end{figure}

At low temperatures, the thermodynamics is dominated by Goldstone phonons.  Using standard results in the phonon regime, one can show (see ~\ref{phononthermosec}) that in the $T\to 0$ limit,
\beq \sqrt{\frac{\rho_{s0}}{\rho_{n0}}\lp x} \simeq \frac{2}{3}\label{smallparameteridentity4}\eeq
for a unitary Fermi gas.  
This non-zero limiting value as $T\to 0$ arises from the fact that \textit{both} $\lp x$ and $\rho_{n0}$ vanish as $T^4$.  
For a Fermi gas at unitarity in the $T\to 0$ limit, using (\ref{smallparameteridentity4}) and the fact that $x=1/3$ (see ~\ref{phononthermosec}), (\ref{vsovervn4}) and (\ref{vsovervn5}) give
\beq   \frac{v^{(1)}_s}{v^{(1)}_n} \simeq \frac{3}{5},\;\;\;\frac{\rho_{s0}v^{(2)}_s}{\rho_{n0}v^{(2)}_n} \simeq -2,\label{vsovervn6}\eeq
even though $\lp$ vanishes as $T\to 0$ (as shown in figure~\ref{lpfig}).  If we had simply set $\lp=0$, as often done in the superfluid literature, we would instead obtain the results given by (\ref{vfields0}).

In figure~\ref{velocityratiofig}, we plot the velocity ratios in (\ref{vsovervn4}) and (\ref{vsovervn5}) for a superfluid unitary Fermi gas.  It is clear that the superfluid and normal fluid velocities which are involved in first and second sound are significantly different from those given in (\ref{vfields0}).  This is a reflection of the fact that first and second sound are not pure uncoupled density and temperature waves, respectively, which would be described by (\ref{vfields0}) and correspond to $\lp=0$ (and $\rho_{n0}\neq 0$).  As we have shown in \cite{tayhu09}, such in-phase ($\bv_s=\bv_n$) and out-of-phase zero current ($\bj=0$) solutions correspond to first and second sound velocities given by
\beq u^2_1 = v^2_{\bar{s}},\;\;\;u^2_2 = T\frac{\bar{s}^2_0}{\bar{c}_v}\frac{\rho_{s0}}{\rho_{n0}}\equiv v^2.\label{cspeeds0}\eeq
As we show in figure~2 of \cite{tayhu09}, these sound velocities are in good agreement with a full calculation including the coupling associated with $\lp$.  In this Appendix and section~\ref{sqomegaweightssec}, we have carried out a careful analysis based on expanding to first order in the parameter $\lp x$, which is small ($\ll 1$) at all temperatures~\cite{vinen70,Hohenberg73}.  The results in figure~\ref{velocityratiofig} show that the main effect of working with $\gamma$ not equal to unity is simply that the second sound speed $u_2$ is given by $v/\sqrt{\gamma}$ as in (\ref{cspeeds}), instead of $v$ in (\ref{cspeeds0}).  In contrast, the first sound speed is always very well approximated by $v_{\bar{s}}$, even when $\gamma$ deviates from unity.  

Initially, it seems surprising that the superfluid and normal fluid velocity fields shown in figure~\ref{velocityratiofig} are so different from those in (\ref{vfields0}), since the first and second sound velocities are fairly well approximated by (\ref{cspeeds0}). (See figure~2 of \cite{tayhu09}.)   This arises because of two features evident in (\ref{vsovervn4}) and (\ref{vsovervn5}).  First of all, we note that the first and second sound speeds in (\ref{uexpansion}) involve corrections of order $\lp x\ll 1$, while the related corrections are much larger in (\ref{vsovervn5}) and (\ref{vsovervn6}) since they enter as $\sqrt{\lp x}$.  The second, and more important, reason is that the corrections involve the factors $\sqrt{\rho_{s0}/\rho_{n0}}$ and $\sqrt{\rho_{n0}/\rho_{s0}}$, which are very large at $T=0$ and $T=T_c$, respectively.  

Substantial deviations from the ratios given in (\ref{vfields0}) also occur in superfluid $^4$He, although the magnitude will be different from the case of a Fermi superfluid at unitarity shown in figure~\ref{velocityratiofig}.  This feature was not commented on in the older superfluid $^4$He literature~\cite{ferrel68,vinen70,Hohenberg73}, which concentrated entirely on the frequency and damping of the first and second sound resonances appearing in the dynamic structure factor.  The associated velocity fields calculated here were not discussed.  These velocity fields may be direct experimental interest in future studies of two-fluid hydrodynamics in Fermi superfluids.    

Calculation shows that the various terms in (\ref{vsovervn4}) and (\ref{vsovervn5}) are not small compared to unity.  However, if one formally expands these expressions to leading order in $b^2/v^2_{\bar{s}}$ $(=\sqrt{\rho_{n0}\lp x/\rho_{s0}})$, one finds 
\beq \frac{v^{(1)}_s}{v^{(1)}_n} = 1- \delta,\;\;\frac{\rho_{s0}v^{(2)}_s}{\rho_{n0}v^{(2)}_n} = -1 - \delta,\label{Lifshitzexpansion}\eeq
where the leading correction to (\ref{vfields0}) is found to be
\beq \delta =\frac{\rho_0}{\rho_{n0}}\frac{b^2}{v^2_{\bar{s}}} =\frac{\rho_0}{\rho_{n0}}\sqrt{\frac{\rho_{n0}}{\rho_{s0}}\lp x}
.\eeq
This value of $\delta$ agrees with the expression originally worked out by Lifshitz (see page~71 in \cite{Khalatnikov}) in the limit where $u^2_2\ll u^2_1$, valid close to $T_c$.  In the case of superfluid $^4$He, the fact that $\lp\ll 1$ even close to $T_c$ (it diverges very weakly at $T_ c$) means that the first correction in (\ref{Lifshitzexpansion}) is indeed small, $\delta\ll 1$.  In contrast, for a Fermi gas at unitarity, $\lp$ is of order unity or larger near $T_c$ (see figure~\ref{lpfig}) and as a result we find $\delta = 0.3$ just below $T_c$~\cite{note}.  In this case, the expansion in (\ref{Lifshitzexpansion}) is not valid, as can be seen by comparing (\ref{Lifshitzexpansion}) with the results in figure~\ref{velocityratiofig}.


\begin{thebibliography}{99}
\bibitem{note} With a better theory of the thermodynamics than NSR, we would find that $\delta$ diverges at $T=T_c$ (as it does in superfluid $^4$He).  Our point here is that it is also significant in a large region below $T_c$.
\end{thebibliography}
\end{document}